# Bulk-like structural, magnetic and optical properties of (111)- and (001)-NiO thin films

S. Kaur[1], Smriti Bhatia[1], Pooja[1], Kshitij Sharma[2], V. K. Malik[3], J. P. Singh[1], K. Sen[1]*

[1]Department of Physics, Indian Institute of Technology Delhi, Hauz Khas, New Delhi 110016, India

[2]Central Research Facility, Indian Institute of Technology Delhi, Hauz Khas, New Delhi 110016, India

[3]Department of Physics, Indian Institute of Technology Roorkee, Roorkee 247667, India

*Email: kaushik.sen@physics.iitd.ac.in

We have grown (111)- and (001)-oriented NiO thin films on (0001)-Sapphire and (001)-MgO substrates using pulsed laser deposition (PLD), respectively. DC magnetic susceptibility measurements underline that the Néel temperatures of the samples are beyond room-temperature. This is further confirmed by the presence of two-magnon Raman scattering modes in these films in ambient conditions. Moreover, relative intensity of the two magnon-mode with respect to a neighboring phonon mode in the films, at least down to 30 nm thickness, is comparable to the same for bulk NiO. UV-vis spectroscopy and spectroscopic ellipsometry determined that the bandgap of the films is 3.6 eV which is well within the range for bulk NiO. Thus, these indicate that the thin films are bulk-like. Further, photoluminescence measurements on (111)-NiO films obtained two-radiative transitions at 385 and 405 nm. The linewidth of the latter broadens towards low temperatures, indicating a plausible exciton-magnon coupling. Overall, these PLD-grown oxide films hold significant technological importance due to their optical transparency and their capacity to host robust magnons at room temperature.

## 1. Introduction

Nickel oxide (NiO) is an antiferromagnetic insulator with a band gap ranging from 3.6 to 4.0 eV and the Néel temperature of 523 K. The antiferromagnetic order in NiO involves ferromagnetic spin alignments within the (111) planes, with these planes being antiferromagnetically coupled along the [111] direction [1]. NiO has a wide range of applications in devices such as spin-valves [2,3], exchange-biased systems [4], and UV light-emitting diodes [5]. However, NiO thin films are prone to develop defects due to nickel vacancies or oxygen present in interstitial sites[6],[7]. Such defects, if present in large amounts, would affect both the magnetic and optical properties of the material. Thus, improved device performances necessitate the use of crystalline films. Traditional deposition techniques, such as pulsed laser deposition (PLD) [8], [9], [10], magnetron sputtering [11], and molecular beam epitaxy [12] have been proven to grow NiO films with reasonable epitaxial quality and smooth surface morphology. The reasonably good quality epitaxial NiO films were grown predominantly on two substrates (0001)-oriented sapphire and (001)-oriented MgO substrates using pulsed laser deposition [9], [13], [14], [15], sol-gel method [16], molecular-beam epitaxy [17] and RF magnetron sputtering [18]. The orientation of NiO thin films on these substrates are (111) and (001), respectively. While (111)-oriented NiO films are found to have twinned domains, (001)-NiO are monodomain. However, the study in which the structural, magnetic, and optical properties of epitaxial NiO films are discussed together is missing [9], [11], [14], [19], [20], [21], [22], [23], [24]. In most reports, these properties are examined individually. For instance, Lee et al. observed that higher substrate temperature enhanced the crystallinity of (111)-NiO deposited on (0001)-sapphire [11]. Yamuchi et al. investigated the structural properties of (111)-NiO epitaxial thin films grown on atomically stepped sapphire (0001) substrates at room temperature using pulsed laser deposition [16]. Nishimoto et al. demonstrated that (001)-NiO on (001)-MgO forms single-phase quadrangular grains, whereas (111)-NiO on (0001)-Sapphire exhibits a double-domain structure with two types of triangular grains rotated by 60° [23].

Feldl et al. reported two-magnon Raman modes at room temperature in (001)-NiO thin films of thickness 60 nm, observing a significantly weaker relative intensity compared to bulk NiO, likely due to point defects affecting the magnetism of the film [20]. Attri et al. reported a high Néel temperature above room temperature in (111)-NiO thin films, but Raman scattering failed to detect the two-magnon mode, likely due to defect-induced suppression [19]. Lahiji et al. utilized spectroscopic ellipsometry to characterize (111)-NiO thin films, identifying a direct bandgap shift from 3.65 eV to 3.85 eV dependent on oxygen content, with the refractive index varying slightly between 2.36 and 2.39 [23]. Lu et al. examined the impact of annealing temperature on the optical properties of NiO thin films on (001)-Si substrates, demonstrating that the optical band gap remains constant at 3.8 eV while the refractive index and film thickness decreases with increasing rapid thermal annealing (RTA) temperature [25]. Further, antiferromagnetic order gets suppressed in thin films compared to bulk NiO [20,26]. In the present work, we offer a comprehensive analysis that addresses all these characteristics collectively.

This manuscript reports the pulsed laser deposition of highly oriented (111)- and (001)-NiO thin films on (0001)-sapphire and (001)-MgO substrates, respectively. X-ray diffraction and atomic force microscopy measurements testify for the high crystallinity and smooth surface of the films. DC magnetic susceptibility data along with Raman scattering signature of two-magnon mode confirm that the NiO films are in bulk-like antiferromagnetic state. Further, UV-vis spectra and spectroscopic ellipsometry data confirm that the optical properties of the films are bulk-like with the bandgap of 3.6 eV. Finally, we present photoluminescence data that identify two radiative transitions when (111)-NiO is excited with the light having energy equal to its bandgap. Temperature-dependent linewidth of one of the transitions indicates a possible exciton-magnon interaction. In summary, we developed (111)- and (001)-NiO thin films that are optically transparent and host robust bulk-like antiferromagnetic state and magnons at room temperature.

## 2. Experiments
### 2.1 Pulsed laser deposition

We grew NiO thin films using pulsed laser deposition (PLD) from a polycrystalline NiO target. We pre-ablated the target at room-temperature before every deposition with the laser pulses of 1500 at 1.1 J/cm$^2$ and frequency of 4 Hz. We grew (111)- and (001)-oriented NiO films on (0001)-sapphire and (001)-MgO substrates, respectively. Prior depositions, substrates were heated to 500 °C and kept at that temperature for 15 minutes in oxygen with the partial pressure of $P_{O_2} = 0.54 \times 10^{-3}$ mbar. The depositions were made in the same $P_{O_2}$ with the laser fluence of 1.1 J/cm$^2$ and a repetition rate of 4 Hz. After depositions, we performed an *in-situ* annealing for 30 min at the same temperature of 500 °C, however, in an increased $P_{O_2}$ of 12 mbar. Subsequently, the samples were cooled down to room temperature at 30 °C/min rate with the continuous reduction of pressure from $P_{O_2} = 12$ to the base pressure of $0.54 \times 10^{-6}$ mbar. We optimized the growth rate with the help of x-ray reflectivity measurements. We determined 5000 pulses are required to grow 30 nm thick films, which yields the growth rate of 0.006 nm per pulse. The present study focusses on thick films that were grown using 15,000 pulses. Since, the growth is not completely linear throughout the entire thickness, we performed cross sectional field emission scanning electron microscopy to estimate the average thickness of ~120 nm (not shown here). Note that, the used growth rate of 0.006 nm per pulse is extremely slow, which leads to high quality thin films with reduced defects.

### 2.2 Structural characterization

We used x-ray diffraction (XRD), x-ray reflectivity (XRR), x-ray reciprocal space map (RSM), cross-sectional field emission scanning electron microscopy (FESEM), and atomic force microscopy (AFM) for structural characterization, including surface morphology of the samples. We used a PANalytical X'Pert PRO diffractometer with Cu-$K_\alpha$ radiation of $\lambda = 1.54$ Å for XRD, XRR and RSM measurements. For FESEM measurement we used Magna LMU Model by TESCAN. For AFM measurements, we used a Bruker-made setup.

### 2.3 Magnetization measurements

We measured DC magnetization ($M$) versus temperature ($T$) of the samples using a Quantum Design made MPMS setup. The samples were mounted in plastic straws in such a way that the applied magnetic field was parallel to the plane of the samples. For $M$-$T$ measurements in zero field cooled (ZFC), we cooled the samples from 300 K to 10 K in zero magnetic field. Subsequently, the data were recorded while warming up to 350 K in the magnetic field of 200 Oe. The field-cooled (FC) data were recorded right after ZFC while cooling down to 10 K under the same magnetic field. We performed corrections for diamagnetism by recording $M$ vs. applied magnetic field of the samples at room temperature. In addition, we measured $M$ vs. $T$ of polycrystalline NiO powder.

### 2.4 Raman scattering

The Raman scattering measurements were carried out using a confocal WiTech setup with a UV laser excitation of 266 nm. The UV laser helps to confine depth-of-focus onto the substrate surface, which drastically reduces Raman response from substrates. All measurements were performed at room temperature in ambient conditions by focusing the laser beam onto the sample surfaces through a microscope objective of 40 ×. We used a grating of 2400 grooves/mm to obtain the Raman response over a wide energy range.

### 2.5 UV-vis spectroscopy and photoluminescence

UV-vis spectra were recorded in transmission mode for the wavelength range of 200 to 800 nm using PerkinElmer LS-55. We removed the substrate contribution by recording transmission spectra of the bare substrates, and then dividing the transmission spectra of the samples by this. Finally, we obtained two similar peaks, one narrow and the other broad, in both (111)-NiO and (001)-NiO, confirming no artifacts in our data analysis.

We obtained photoluminescence (PL) spectra of (111)-NiO films using a Horiba-made Quanta Master steady-state spectrometer. We chose an excitation wavelength of 345 nm which equals the band gap of the film. Temperature-dependent PL data for a (111)-NiO film and a bare sapphire substrate were recorded by gluing them with GE varnish onto a copper strip attached to the cold finger of the cryostat. We found a total of seven and five Lorentzian peaks were necessary to fit PL response of the sample and the substrate, respectively. The linewidth and position of the five out of seven peaks in the sample were the same as of the five peaks of the substrate. Further, the background in the model for both the sample and the substrate were assumed to be the same. Subsequently, we extracted PL response of only the NiO film by subtracting these five peaks from the entire fitting of the PL response of the sample.

### 2.6 Optical ellipsometry

The refractive index and extinction coefficient of (001)- and (111)-NiO thin films were obtained through the spectroscopic ellipsometry method. A Woollam Ellipsometer AlphaSE was used to record spectra between 200 and 1700 nm. The measurement angle was 65°.

We extracted thin film responses using the two-layer model, i.e. substrate/NiO. The first step involves the measurements of ellipsometry angle (Ψ) and phase difference (Δ) of the bare (0001)-Sapphire and (001)-MgO substrates. Subsequently, we measured the same for (111)- and (001)-NiO thin films. The optical properties of NiO films on both substrates were analyzed using a general oscillator model. We collected spectroscopic ellipsometry (SE) data for the bare substrates to exclude the contribution of the substrates from the thin film samples. The SE data for bare substrates were fitted with Cauchy function model. We used the Cody-Lorentz oscillator which is a more generalized feature of the Lorentz oscillator model for (111)-NiO. For (001)-NiO, we have used the Gaussian oscillator model.

### 3. Results and discussion
### 3.1 Structural properties of NiO thin films

We grew (111)- and (001)-oriented NiO films of ~120 nm thicknesses on (0001)-oriented (hexagonal notation) sapphire ($Al_2O_3$) and (001)-

oriented MgO substrates, respectively. Our choice of substrates for the preferred orientations of NiO films stem from the existing reports[27],

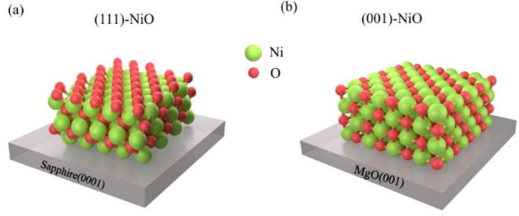

**Figure 1**. Sketches illustrating (111)- and (001)-oriented crystallographic NiO layers on (a) sapphire and (b) MgO substrates, respectively.

[28], [29], [30], [31],[32]. Figure 1(a) and Figure 1(b) show (111)- and (001)-oriented crystallographic NiO layers.

In (111)-NiO, the lattice planes parallel to the substrate surface consist of either Ni or O atoms. Since sapphire is an oxide, the very first layer in (111)-NiO consists of Ni atoms. However, in (001)-NiO, the lattice planes parallel to the substrate surface have a mixture of Ni and O atoms.

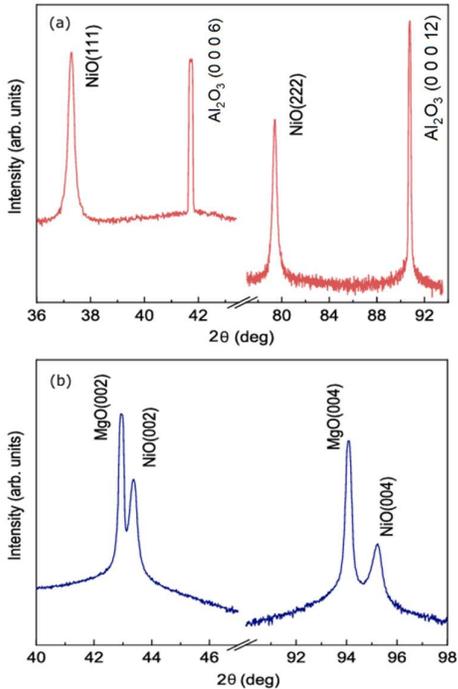

**Figure 2**. XRD data using symmetric $2\theta$-$\omega$ scans for (a) (111)-NiO and (b) (001)-NiO films

Considering $a$-axis lattice parameters of $a_{NiO} = 4.177$ Å and $a_{Sapphire} = 4.758$ Å, the in-plane lattice mismatch between (0001)-Sapphire and (111)-NiO yields:

$$\frac{\sqrt{2}a_{NiO} - a_{Sapphire}}{a_{Sapphire}} \times 100 = 24.15\,\%.$$

However, a domain matching considering the symmetry of the (0001)-Sapphire and (111)-NiO crystallographic layers reduces the lattice mismatch to 7.52 % [11]. Following ref. [11], the mismatch is quantified as follows:

$$\frac{(\sqrt{2}a_{NiO}\sin 60^o) \times 2 - a_{Sapphire} \times 2}{a_{Sapphire} \times 2} \times 100 = 7.52\,\%.$$

For (001)-oriented NiO on (001)-MgO substrate, the trivial lattice mismatch is given by the following equation:

$$\frac{a_{NiO} - a_{MgO}}{a_{MgO}} \times 100 = 0.831\,\%,$$

where $a_{NiO}$ and $a_{MgO}$ are 4.177 Å and 4.212 Å, respectively. In both cases, such mismatches give rise to compressive strains to the films. We further summarize this lattice mismatch in

Table **1**.

**Table 1**. Underlying lattice mismatches between the oriented substrates and NiO films. Here, we consider bulk lattice parameters.

| Orientation | Mismatch (%) |
|---|---|
| (111)-NiO on sapphire | 7.52 [11] |
| (001)-NiO on MgO | 0.831 [20] |

We determined the epitaxial nature of the NiO thin films by performing XRD measurements in symmetric $2\theta - \omega$ mode. Figure 2 (a) and Figure 2(b) show XRD data of NiO films grown on (0001)-sapphire and (001)-MgO substrates, respectively. In case of NiO/sapphire, the occurrence of only (111) and (222) Bragg peaks of NiO testify for the highly oriented NiO films with the surface normal along [111]. We observed only (002) and (004) Bragg peaks of NiO in the case of NiO/MgO. This confirms that the NiO films grown on MgO substrates are (001)-oriented. Further, comparable Bragg peak linewidths of the single-crystal substrates and the NiO films suggest high crystalline quality of the films.

We determined lattice parameters of the films from the (111) and (002) Bragg peaks by considering the cubic symmetry of NiO, as shown in Table 2. The out-of-plane lattice parameters are 4.168 Å for (111)-NiO and 4.180 Å for (001)-NiO, whereas the bulk lattice parameter for cubic NiO is 4.176 Å.

Considering cubic structure of NiO (rock salt structure, $Fm\bar{3}m$ space group), the resulting lattice deviations for (111)- and (001)-NiO films with respect to bulk NiO are 0.192 % and 0.096 %, respectively.

**Table 2**. Lattice parameters ($a_{pc}^{exp}$) of pseudocubic unit cells in (111)- and (001)-oriented NiO films. Lattice parameter of bulk NiO is also mentioned.

|  | $a_{pc}^{exp}$ (Å) | $a^{bulk}$ (Å) | Deviation (%) |
| --- | --- | --- | --- |
| (111)-NiO | 4.168(1) | 4.176 | 0.192 |
| (001)-NiO | 4.180(1) |  | 0.096 |

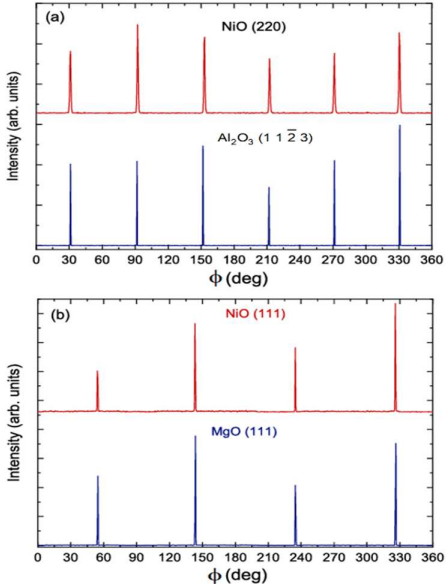

**Figure 3.** $\phi$-scan around (a) (220) Bragg plane of NiO and around (1 1 $\bar{2}$ 3) Bragg plane of Al$_2$O$_3$ for (111)-NiO, and (b) (111) Bragg plane of NiO and MgO for (001)-NiO.

The orientation of in plane crystalline domains for (111)-NiO and (001)-NiO thin films was examined using high resolution XRD $\phi$-scans. Figure 3(a) presents the φ-scans for the NiO (220) and sapphire (1 1 $\bar{2}$ 3) reflections. There are six peaks observed for sapphire (1 1 $\bar{2}$ 3) reflection, which is in agreement with the fact that it has a hexagonal crystal structure [22]. The NiO (220) reflection displayed a six-fold rotational symmetry instead of the three-fold rotational symmetry expected for a single (111)-oriented NiO film, suggesting that the NiO film grown on the (0001) sapphire substrate exhibits two distinct in-plane orientations. These orientations indicate the presence of twin domains, with each domain rotated by 60° relative to the other, replicating the hexagonal symmetry of the (0001)-sapphire substrate. This twinning behaviour aligns with the structural relationship between the cubic NiO film and the hexagonal sapphire, resulting in the observed six-fold symmetry. Such a result is commonly reported for (111)-oriented NiO films on (0001) sapphire substrates [9], [10], [11], [22].

In comparison, Figure 3 (b) presents φ-scans for the NiO (111) and MgO (111) reflections for (001)-NiO film. The NiO (111) reflection on the (100) MgO substrate exhibited a four-fold rotational symmetry, indicating that the NiO film on the (100) MgO substrate has a single in-plane orientation. This observation is consistent with the expected symmetry for (100)-oriented NiO films, which naturally exhibit four-fold symmetry and a single-domain in-plane orientation [33]. The substrate peaks align perfectly with the layer peaks, indicating that the crystallographic directions of the substrate and the layer are in perfect alignment. This confirms a cube-on-cube epitaxial relationship between the substrate and the layer [12], [16],[22], [34].

Figure 4 (a) shows reciprocal space map around (0 0 0 6) Bragg peak of sapphire for the (111)-NiO films grown on sapphire substrates. The broadening in the $Q_{\parallel}$ gives the information about mosaic texture in the film. The (111) Bragg peak of NiO appears at the expected position for corresponding bulk NiO, as indicated in figure 4(a). Further, the Bragg peak almost symmetrically distributed around the bulk position, and clearly shifted towards left with respect to the vertical line that connects all (0 0 0 *l*) peaks of sapphire. This indicates that *c*-axis lattice parameter of the bulk NiO and the (111)-NiO film is nearly the same. Thus, ~120 nm thick (111)-NiO films is almost-fully fully relaxed.

Figure 4 (b) shows the reciprocal space map around (024) Bragg peak of MgO for (001)-NiO film grown on MgO substrate. The vertical dotted line connects all the (02$l$)-peaks of MgO. The (024) Bragg peak of NiO is asymmetrically distributed around the vertical line. Specifically, the diffraction intensity is shifted towards the bulk NiO (024) peak, which indicates a partial strain relaxation in ~120 nm thick films. In summary, ~120 nm thick (111)- NiO are almost-fully strain-relaxed, whereas (001)-NiO is partially strain-relaxed. This is the case, as the lattice mismatch between substrate and film is large for (0001)-sapphire in comparison to (001)-MgO, as given in

**Table 1**.

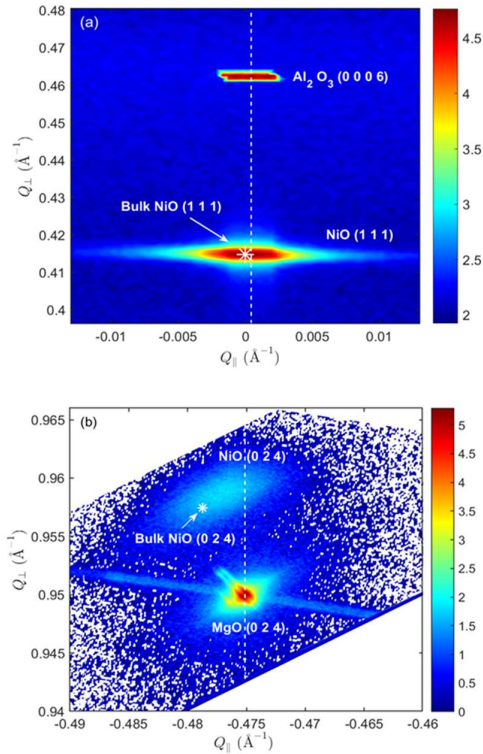

**Figure 4**. Reciprocal space map around (a) (0 0 0 6) peak of sapphire for (111)-NiO and (b) (024) peak of MgO for (001)-NiO. The dashed white line connects the (0 0 0 $l$) – peaks of Sapphire for (111)-NiO and (02$l$) peaks of (001)-NiO. The asterisks show the location of the corresponding Bragg peaks for bulk NiO.

Figure 5 (a) and Figure 5 (b) show AFM images of (111)- and (001)-NiO thin films over an area of 200 nm × 200 nm. We obtained an average root-mean-square surface roughness of 0.8419 nm and 0.2649 nm, respectively. These values correspond to two unit-cells and half unit-cell of NiO, respectively. Thus, with such reasonably small surface roughness in the films along with their epitaxial nature, we can infer that the ~120 nm thick (111)-NiO and (001)-NiO films are of high-quality.

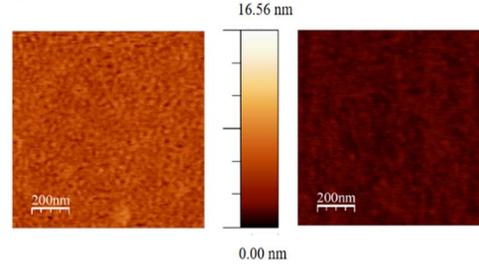

**Figure 5**. AFM images of (a) (111)-NiO and (b) (001)-NiO, respectively.

### 3.2 Magnetic properties of the films

Figure 6 (a) and Figure 6 (b) show DC magnetic susceptibility ($\chi$) of polycrystalline NiO powder and (111)-NiO thin films as a function of temperature ($T$), respectively. The measurements were carried out by applying the magnetic field of 200 Oe. In the case of films, the magnetic field is applied parallel to the sample surface. We didn't observe any clear transition from paramagnetic state to antiferromagnetic state in the measurement range of 10 to 325 K both for bulk and film NiO. Our measurements do not allow to determine the exact Néel temperature ($T_N$) of the films. However, we can conclude that the films are antiferromagnetic at least below room temperature [21], and $T_N$ of the films are beyond 325 K. For bulk NiO, $T_N = 523$ K [35]. Here, we did not show $\chi$ vs. $T$ for (001)-NiO, as large paramagnetic contribution from MgO substrates dominate the data.

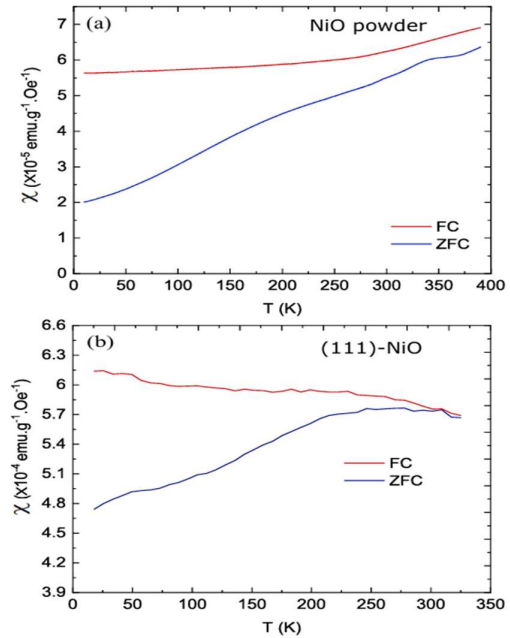

**Figure 6**. DC magnetic susceptibility ($\chi$) versus temperature (T) for (a) polycrystalline NiO powder and (b) (111)-NiO films, respectively.

### 3.3 Phonons and magnons in (111)- and (001)-NiO films

We do not expect first-order phonon modes in bulk NiO due to its cubic rock-salt structure. However, due to crystal imperfection, a group of first-order phonon modes (1P) appear in polycrystalline NiO powder between 400 and 600 cm$^{-1}$, as shown in Figure 7. Such 1P modes are made of both longitudinal optic (LO) and transverse optic (TO) modes. The next three bands between 600 and 1200 cm$^{-1}$ are second-order phonon modes. Two TO modes appear at ~720 cm$^{-1}$; a combination of TO and LO modes appears at ~894 cm$^{-1}$; and two LO modes appear at ~1084 cm$^{-1}$. Finally, the strongest mode at 1490 cm$^{-1}$ arises from two-magnon (2M) Raman scattering involving excitations of any two magnons of opposite momenta near zone boundary [36]. Effectively, this 2M mode represents magnon density of states for the antiferromagnetic ground state of bulk NiO (polycrystalline powder) at room temperature.

In addition to Raman response from bulk-NiO, Figure 7 shows Raman scattering data from 120 nm (001)-NiO, 120 nm (111)-NiO, and 30 nm (111)-NiO films. A prominent difference between the bulk and thin films is the drastically diminished 1P mode between 400 and 600 cm$^{-1}$ Raman data reaffirm that our thin films of (111) and (001) orientations are highly crystalline in films. As pointed out earlier, the origin of this 1P mode is crystal imperfection. Thus, our Raman data reaffirm that our thin films of (111) and (001) orientations are highly crystalline.

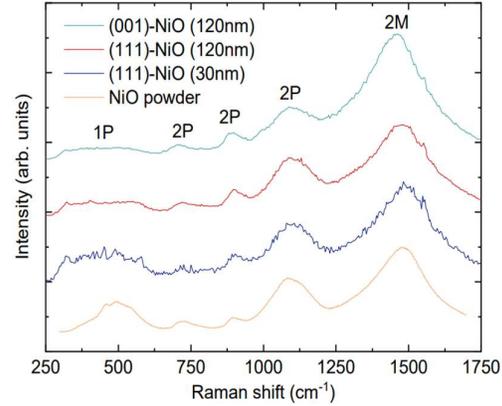

**Figure 7**. Raman spectra of 120 nm thick (001)-NiO, 120 and 30 nm thick (111)-NiO films. In addition, Raman spectrum of polycrystalline NiO is given. All measurements were carried out at room temperature.

To understand the robustness of the antiferromagnetic ground state in thin films, we introduce a factor $F_Q = \frac{I_{2M}}{I_{2P}}$ which is the ratio of the intensity of the 2M mode and the neighboring 2P mode. For bulk NiO, $T_N = 523$ K, and $F_Q$ is 2.38. It is expected that $T_N$ would decrease in thin films, and so the factor $F_Q$. However, we can maintain $F_Q$ at the same level of bulk at room-temperature in 120 nm thick (111)- and (001)-NiO films, as shown in Figure 8. This indicates that the antiferromagnetic order in our films is as strong as in the bulk, and $T_N$ in films is beyond room temperature. In 30 nm thick (111)-NiO, we observed similar $F_Q$, which attesting that bulk-like antiferromagnetic state persists down to at least 30 nm thick NiO films.

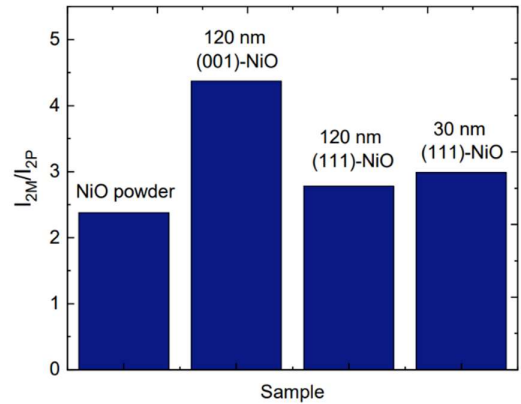

**Figure 8**. Ratio of the intensity of 2M peak to the intensity of 2P peak for NiO powder, 120nm (111)-NiO, 120nm (001)-NiO, and 30nm (111)-NiO.

The appearance of a robust antiferromagnetic state in the thin films indicates a reduced number of defects compared to films reported in previous studies, such as those grown via plasma-assisted molecular beam epitaxy [21] and plasma-enhanced atomic layer deposition techniques [19]. In the first reference, the 2M mode was weaker in (001)-oriented NiO thin films compared to bulk NiO, while in the second reference, no 2M mode was detected in (111)-oriented NiO films. Although comparing films grown using different methods is challenging, we believe two critical growth parameters significantly improved the film quality by reducing defects. The first is the slow growth rate of only 0.006 nm/pulse. The second is the 30-minute annealing process in a significantly high oxygen partial pressure of 12 mbar, in contrast to the PLD growth conditions reported in ref. [9].

As shown in Figure 7, all 2P and 2M modes in (111)-NiO films match with the corresponding positions in bulk NiO. This attests that the lattice dynamics and spin dynamics in (111)-NiO films and bulk-NiO are identical. In the case of (001)-NiO, the 2P modes largely match with the bulk NiO. However, its 2M mode shows a sizeable softening 20 cm$^{-1}$. The origin of such softening is unlikely to be due to strain effects, as strain would also affect the 2P modes. We believe that the different orientations of NiO affect the effective magnetic anisotropy in the films, thereby influencing the exchange interactions and consequently the energy scale of the 2M mode.

### 3.4 Optical bandgaps and refractive indices of (111)- and (001)-NiO

Figure 9 (a) shows transmission spectra of 120 nm thick (111)- and (001)-oriented NiO thin films. Two peak-like features are attested at 560 nm (significantly broad) and at 360 nm (narrow). In general, the spectra are very similar which indicates no impact of substrate-induced epitaxial strains on the optical properties of 120 nm thick films. A sharp drop in transmission is observed below ~360 nm, which is indicating of the electronic transition across the optical band gap [37,38]. We precisely determined the bandgap by taking derivatives of these transmission spectra [39], as shown in Figure 9 (b). In both cases, the peaks in derivatives appear at 345 nm which is 3.59 eV in energy unit. Thus, the optical band gaps of $E_g = 3.59$ eV are the same for both films. The same bandgap values once again indicate that the films are strain-relaxed, as expected for films of 120 nm thicknesses. Note that this value of the band gap falls within the range of 3.6-4.0 eV, which corresponds to the bandgap of bulk NiO [40]. In the following, we determine the nature of the bandgap, i.e., whether it is direct or indirect bandgap.

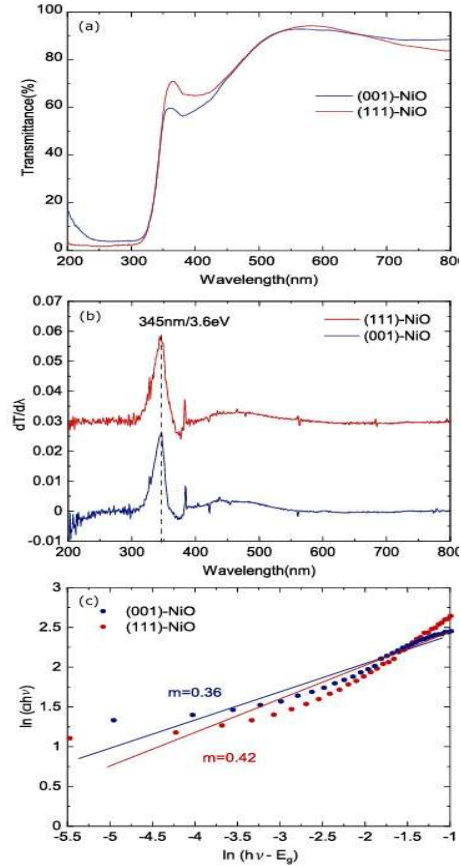

**Figure 9.** (a) UV-vis transmission spectra of (111)- and (001)-NiO thin films of 120 nm thickness. (b) The first derivative of the corresponding transmission spectra highlighting bandgaps. (c) ln(αhυ) vs. ln(hυ-E$_g$). The terms are described in text.

Frequency dependent optical absorption coefficient ($\alpha$) is linked to $E_g$ by the following equation.

$$\alpha = \frac{A}{h\nu}(h\nu - E_g)^m \tag{1}$$

, where $A$ is a material-dependent constant, $h$ is the Planck constant, $\nu$ is the frequency of the incoming radiation, and the value $m$ represents the nature of electronic transition. In particular, $m = 1/2$ and 2 are for allowed direct and indirect transitions, respectively. $m = 3/2$ and 3 are for forbidden direct and indirect transitions, respectively [23,39,41]. Eq. (1) indicates a straight line if $\ln(\alpha h\nu)$ is plotted as a function of $\ln(h\nu - E_g)$. The slope of this straight line yields the value of $m$. For straight line fitting, we used $E_g = 3.59$ eV, as obtained from Figure 9 (b). Figure 9 (c) shows $\ln(\alpha h\nu)$ vs. $\ln(h\nu - E_g)$ plots along with the straight-line fits. We found $m$ values of 0.42 and 0.36 for (111)-NiO and (001)-NiO, respectively, are close to 0.5, indicating allowed direct transitions [41].

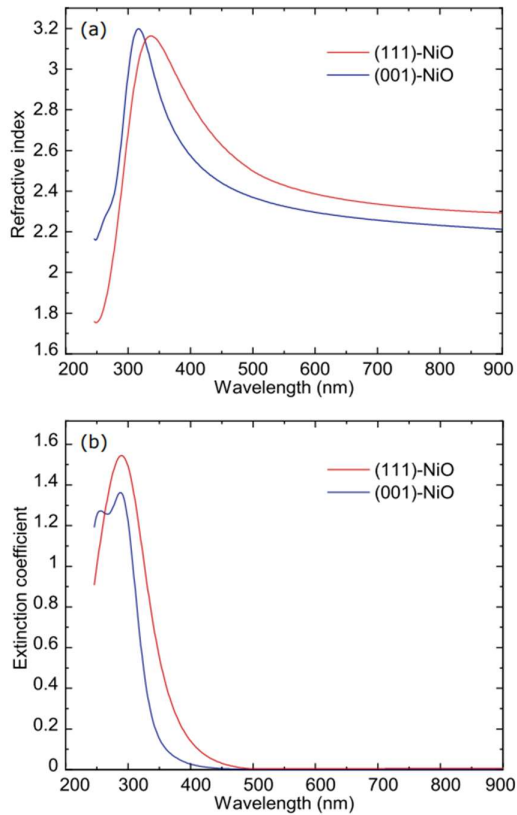

**Figure 10.** (a) Refractive indices and (b) extinction coefficients as a function of wavelength for both (111)-NiO and (001)-NiO films.

We further determined refractive indices ($n$) and extinction coefficients ($k$) of the films using optical ellipsometry. Figure 10 (a) and Figure 10 (b) show $n$ and $k$ as a function of wavelength ($\lambda$) for both (111)- and (001)-NiO films, respectively. The dependence of $n$ and $k$ on wavelength is similar to the previously reported data for NiO thin films and crystals [25,42,43]. The sharp change in $n$ and $k$ around 325 cm$^{-1}$ is associated with the electronic band gaps of the materials.

### 3.5 Photoluminescence spectroscopy of (111)-NiO

We studied dynamics of bound charges in (111)-oriented NiO thin films using temperature dependent photoluminescence measurements within the temperature range of 10 to 400 K. Figure 11 shows PL spectra at several temperatures for the excitation wavelength of 345 nm that corresponds to the bandgap energy. As shown in Figure 11, PL spectra mainly consist of two peaks. One is at 385 nm falling within the UV and the other one is at 407 nm touching the blue side of the visible range. Near room-temperature, the intensity of both peaks is similar. Overall, their intensity increases with decreasing temperatures, though the intensity of the UV-peak increases more than the other one towards reduced temperatures.

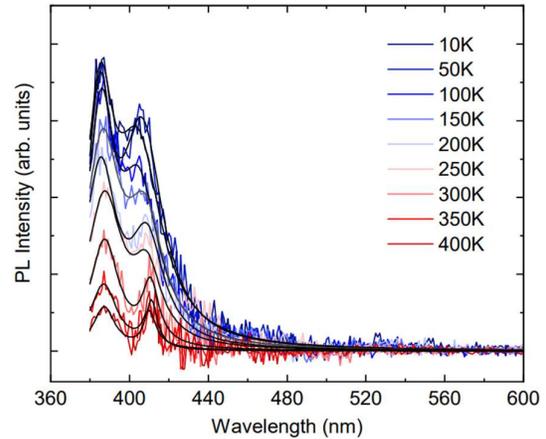

**Figure 11.** PL spectra of 120 nm (111)-NiO at several temperatures. The solid lines through the spectra are the corresponding profile fitting.

At higher temperatures, larger density of photons leads to non-radiative transitions, which thereby decreases the number of emission photons of these two wavelengths. On the other hand, reduced carrier mobility at low temperatures lead to trapping of carriers in shallow states within the bandgap region allow for radiative recombination of electrons and holes [44]. Such temperature dependence of the total PL spectral weight can be modelled by the following Arrhenius equation [45], as shown in Figure 12(c).

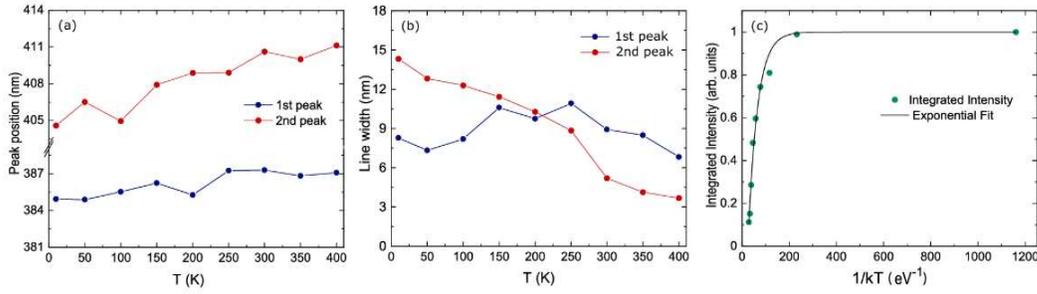

**Figure 12.** (a) Position of the two PL peaks as a function of temperature. (b) Temperature dependence of their linewidths. (c) Integrated PL intensity is plotted as a function of $1/(k_B T)$. The solid line through the data is the Arrhenius fitting.

$$I(T) = I_0 + C \exp.\left(E_a/k_B T\right) \quad (2)$$

, where $I_0$ is the total spectral weight at 0 K, $E_a$ is activation energy of the thermal quenching process, $k_B$ is the Boltzmann constant, $T$ is temperature and $C$ is a constant.

The positions of the PL peaks are also temperature dependent, as shown in Figure 12 (a). The shifting of the PL peaks to higher energy (lower wavelength) towards low temperatures is related to the increase in the band gap [46].

The linewidth of the first peak is nearly temperature independent, shown in Figure 12 (b). Whereas the linewidth of the second peak gets broader with decreasing temperature. Such broadening towards low-$T$ can be understood through exciton interactions with magnetic excitations in the films.

Finally, we come to the origin of the two emission lines. Due to their comparable intensity at room temperature, we believe that the emission-related transitions take place from different excited states to the common valence band states, as shown in Figure 13. Since the bandgap energy is higher than both emission lines, we can assign the UV emission line at ~385 nm to a near band edge emission (NBE) that occurs from recombination of excitons. This emission line, slightly at different wavelength of 346 nm, is previously reported for NiO nanoparticles [47]. The emission line at 405 nm reportedly arises from a defects-induced intermediate state lying within the bandgap. Such defects can be oxygen vacancies or the presence of Ni interstitials [44,48]. Nevertheless, the excited electrons first reach the defects-induced intermediate state through non-radiative transition. The emission occurs because of electronic transition from this intermediate state to the valence band.

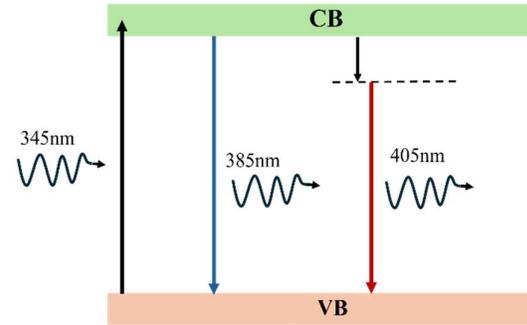

**Figure 13.** Schematic diagram highlighting the origin of two emission lines in PL spectra.

4. **Summary**

In summary, we have reported the growth of (111)- and (001)-oriented NiO thin films using pulsed laser deposition with high crystalline quality, which retain bulk-like physical properties down to at least 30 nm thickness. DC magnetic susceptibility data show that these ~120 nm thick films have Néel temperatures above room temperature, confirming robust antiferromagnetic order at room temperature. Raman scattering responses from the films confirm their room-temperature antiferromagnetic state with the occurrence of a prominent two-magnon mode down to a thickness of 30 nm. Moreover, the relative intensity of this two-magnon mode with respect to a neighboring phonon mode underscores the bulk-like antiferromagnetic state in the thin films. Additionally, our UV-vis spectra, along with the refractive indices and extinction coefficients of the films, demonstrate that their optical properties are bulk-like making the films suitable for optoelectronic applications. Finally, temperature-dependent photoluminescence measurements reveal two radiative transitions and suggest a possible exciton-magnon interaction suggesting the films could be used to study coupling phenomena between electronic and magnetic excitations at room temperature, relevant for magnonic devices and spintronics.

Overall, we have developed high-quality (111)- and (001)-oriented NiO thin films that are transparent and host magnons at room temperature. Thus, the films open opportunities to observe novel magnon properties, such as magnon transport at room temperature.

**Acknowledgement**

This work is supported by the YSRP grant from BRNS, INSPIRE Faculty Fellowship from DST, and CRG from SERB. S. Kaur acknowledges her PhD fellowship from the UGC. All authors acknowledge the experimental support from NRF and CRF at IIT Delhi. All authors thank Sujit Das, Surajit Saha and Malti Kumari for scientific discussions during the development of the manuscript.

---